# Time series segmentation for recognition of epileptiform patterns recorded via Microelectrode Arrays in vitro

Gabriel Galeote-Checa [1], Gabriella Panuccio [4], Angel Canal-Alonso [3], Teresa Serrano-Gotarredona [1], Bernabe Linares Barranco[1],

**1** Instituto de Microelectrónica de Sevilla-IMSE, Centro Nacional de Microelectrónica, Consejo Superior de Investigaciones Científicas and Universidad de Sevilla, Sevilla, Spain
**2** Enhanced Regenerative Medicine Lab, Instituto Italiano di Tecnologia, Genoa, Italy
**3** Instituto de Investigación Biomédica de Salamanca-IBSAL, Consejo Superior de Investigaciones Científicas and University of Salamanca, Salamanca, Spain
**4** Istituto Italiano di Tecnologia, Genoa, Italy

These authors contributed equally to this work.

* ggaleote@ieee.org

## Abstract

Epilepsy is a prevalent neurological disorder that affects approximately 1% of the global population. Approximately 30-40% of patients do not respond to pharmacological treatment, leading to a significant negative impact on their quality of life. Closed-loop deep brain stimulation (DBS) is a promising treatment for individuals who do not respond to medical therapy. To achieve effective seizure control, algorithms play an important role in identifying relevant electrographic biomarkers from local field potentials (LFPs) to determine the optimal stimulation timing. In this regard, the detection and classification of events from ongoing brain activity, while achieving low power consumption through computationally inexpensive implementations, represents a major challenge in the field. To address this challenge, we here present two algorithms, the ZdensityRODE and the AMPDE, for identifying relevant events from LFPs by utilizing time series segmentation (TSS), which involves extracting different levels of information from the LFP and relevant events from it. The algorithms were validated validated against epileptiform activity induced by 4-aminopyridine in mouse hippocampus-cortex (CTX) slices and recorded via microelectrode array, as a case study. The ZdensityRODE algorithm showcased a precision and recall of 93% for ictal event detection and 42% precision for interictal event detection, while the AMPDE algorithm attained a precision of 96% and recall of 90% for ictal event detection and 54% precision for interictal event detection. While initially trained specifically for detecting ictal activity, these algorithms can be fine-tuned for improved interictal detection, aiming at seizure prediction. Our results suggest that these algorithms can effectively capture epileptiform activity, supporting seizure detection and, possibly, seizure prediction and control. This opens the opportunity to design new algorithms based on this approach for closed-loop stimulation devices using more elaborate decisions and more accurate clinical guidelines.

## Author summary

Gabriel Galeote-Checa (Student Member, IEEE) is currently pursuing his PhD in the Neuromorphic Systems research group at the Seville Microelectronics Institute (IMSE-CNM-CSIC), Seville, Spain, part of the Spanish National Research Council (CSIC) and the University of Seville (US). He is also involved in the EU H2020 FET-Proactive "HERMES" project, where he developed a flexible brain implantable device to treat epilepsy. His current research interest includes developing signal-processing techniques for the detection and treatment of seizures via implantable stimulating devices. He is also working at Analog Devices Inc. as a Biomedical Digital Signal Processing engineer.

## Introduction

Epilepsy is a prevalent neurological disorder that affects approximately 1% of the global population [1]. The standard treatment of epilepsy relies on medical therapy; however, 30-40% of the patients respond poorly, if at all, to anti-seizure medications, falling into Drug-Resistant Epilepsy (DRE) patients. Surgical removal of the epileptic focus is the current gold standard for those DRE patients. However, the success of ablative surgery highly depends on the accurate identification of the seizure focus. Brain stimulation has emerged as a promising and less radical alternative to ablative neurosurgery. In recent years, there has been a rapid growth of brain implantable devices, driven by the recent contributions in wireless power transmission techniques [2–4], flexible electronics for implantable devices [3, 5], and device design miniaturization [6–8]. Among the diverse devices under pre-clinical and clinical research, NeuroPace (Mountain View, CA), which is the first FDA-approved responsive neurostimulation system for focal epilepsy, offers a median seizure reduction range of 50-70% across different studies [9]. Given the promising results in seizure suppression by responsive brain stimulation and the latest advances in brain implantable devices, the growth of these devices for treating epilepsy is expected to accelerate in the next years. However, important aspects such as long-term viability, biocompatibility, power harvesting, and the need for more efficient and optimized algorithms make clinical results still improvable.

To operate effectively, brain stimulation devices should detect relevant electrographic biomarkers to determine the optimal stimulation timing and pattern to suppress, or ideally prevent, the seizure. Typically, local field potential (LFP) recordings are used for this purpose. Various biomarkers of interest, such as high-frequency oscillations (HFOs) [10–12], can be detected using embedded algorithms in the implantable device, allowing low latency real-time seizure suppression. In essence, the identification of those different biomarkers in the LFP would help to create decision models for closed-loop brain stimulation. However, the identification of electrographic biomarkers remains challenging due to their diverse characteristics in terms of morphology, time/frequency features, and their correlation with seizure events [12]. To address this challenge, various algorithms have been proposed, such as principal component analysis [13], linear support vector machine [14], approximate entropy [15], K-nearest neighbor and naïve Bayes classification [14], feature extraction using support vector machines [16], or complex multi-feature-based decision models [17]. These algorithms are complex and computationally intensive, and as such their potential for real-time operation is limited. More power-efficient approaches such as phase synchronization in multichannel recordings [18], multilevel wavelet transform [19], Izhikevich neural networks [20], auto-ranging threshold detection [21], or the Wiener algorithm [22], have proven more effective real-time seizure detection in hardware implementations.; However, the main drawbacks are associated with energy and memory consumption, as well as reduced performance caused by simpler implementations due to hardware constraints.

Designing efficient algorithms requires reducing three main aspects: computation, memory, and power consumption. By reducing the computational cost, we can also decrease power consumption. This involves: (i) reducing algorithm complexity by optimizing the number and type of operations performed by it, (ii) minimizing memory access, (iii) increasing parallelism, and (iv) optimizing hardware architecture. Following these principles, we propose two novel algorithms that focus on optimizing computation and memory. Based on the principles of optimization, we have developed two innovative algorithms that aim to reduce computational complexity by minimizing the number of operations required and lowering memory usage. Additionally, our algorithms address the issue of complex and recursive computations, which can be inefficient and time-consuming. By prioritizing efficiency and effectiveness, we hope to create algorithms that can be easily implemented and utilized in a variety of settings. A time series segmentation approach is used to detect and classify epileptiform events within LFPs. This method involves dividing a continuous signal into discrete segments that share common characteristics. The goal is to identify common patterns, states, or behaviors within the data, which will provide varying levels of information. [23, 24]. In the scope of this work, time series segmentation can be used to divide LFPs into distinct categories such as spiking events, firing patterns, or different network states. This approach can help manage a large amount of data and facilitate the application of various algorithms to post-process these segments, such as seizure prediction or control. The objective of this work is to classify ictal and interictal events in LFP data based on their temporal and morphological characteristics. We also aim to determine the current state of the brain network in real time. The principal goal is to ensure that the algorithms we propose have low computational complexity, thereby making them suitable for hardware implementation. We have achieved this goal by developing two new algorithms and validating them through microelectrode array (MEA) recordings of epileptiform activity generated by mouse hippocampus-cortex (CTX) slices treated with the convulsant drug 4-aminopyridine (4AP). This model is widely used to study limbic ictogenesis in vitro. [25], as it offers several parallelisms with the most common type of DRE in humans, i.e., mesial temporal lobe epilepsy (MTLE), including the primary origin of ictal activity in the entorhinal cortex [26–29].

The two algorithms use different approaches; the first (ZdensityRODE) utilizes an adaptive z-score method, which uses a short-term memory strategy to provide a smooth detection of the brain's states. The second algorithm (AMPDE) uses scalogram-based peak detection and density estimation in the signal for extracting events of interest determined by their density of excitation. Both algorithms incorporate a post-processing technique with a cumulative look-forward time integration based on their density and can distinguish among ictal discharges, interictal events, and baseline. The ZdensityRODE algorithm was designed for faster computation and reduced operations, while the AMPDE was designed for better adaptability to long-term variations but at the cost of more complex computation.

## Materials and Methods

### Microelectrode Array recording of epileptiform activity generated by brain slices

Horizontal hippocampus-cortex (CTX) slices (400 *µm*-thick) were obtained from 4-8 weeks old male CD1 mice. Epileptiform activity was acutely induced by treatment with 4-aminopyridine (4AP, 250 $\mu M$). Extracellular field potentials were acquired using a 6 × 10 planar MEA with TiN electrodes (diameter: 30 *µm*, inter-electrode distance: 500 *µm*) and a MEA1060 amplifier. Signals were sampled at 2 kHz, low-pass filtered at half the sampling frequency before digitization, and recorded to the computer's hard drive via the McRack software. All recordings were performed at 32°C. The full methodological details describing brain slice preparation and maintenance can be found in [30]. Animal procedures were conducted in accordance with the National Legislation (D.Lgs. 26/2014) and the European Directive 2010/63/EU, and approved by the Institutional Ethics Committee of Istituto Italiano di Tecnologia and by the Italian Ministry of Health (protocol code 860/215-PR, approval date 24/08/2015). Animals were monitored daily and euthanized under deep isoflurane anesthesia, verified as the lack of reflexes upon feet, paws, and tail pinch, compliant with FELASA standards. The equipment for MEA recording and temperature control was obtained from Multichannel Systems (MCS), Germany. Fig. 1 highlights the electrode mapping relative to the brain slice position on the MEA, showing the different regions from where the LFP signals are recorded.

### Dataset preparation

A validation dataset was generated from 7 brain slices. Brain slices coupled to 60-channel MEAs provided 11 ± 5 (*mean ± SD*) valid electrodes. A total of 68 records, each being 30 minutes long, were extracted and then annotated. Epileptiform events were identified through a semi-automated process. Namely, an automated wavelet-based algorithm was first used to detect the events; then, the marked events were visually inspected and confirmed or manually corrected as required. This is detailed in [31]. Overall, the used dataset comprises a total of 521 ictal events and 6318 interictal events. LFPs were labeled as ictal, interictal, and baseline, coded as 2, 1, and 0, respectively. To ensure the homogeneity of the dataset, we established the following inclusion criteria: i) uniform sampling rate of 2 kHz to standardize disparate sample rates and ensure consistent recording precision; ii) records from the electrodes within the parahippocampal cortex (See Fig. 1). All data and metadata are publicly available via the zenodo repository (10.5281/zenodo.10154332).

### Experimental design: Parameter Search and Optimization Procedure

All records were maintained for training in different scenarios, similar to online operation, under low and high SNR conditions, to avoid a possible overfitting of the algorithm. Table 1 summarizes the dataset characteristics for this experiment. Algorithms were trained with an optimization matrix including variations for each of the parameters of the algorithms. The complete set of combinations is described in Table 2, where the parameter search frame is detailed. The experimental protocol devised for this study is depicted in Figure 2. The dataset is divided into 70-30 test-validation splits. After running the algorithms in the training dataset, the best 10 scores were averaged. This particular number was chosen heuristically to obtain a winning combination. For recordings with low Signal-to-Noise Ratios (SNR) (< 20 dB), parameters were further fine-tuned before the validation stage, to improve the accuracy.

**Table 1. Brain slice MEA electrophysiology dataset description.**

| Parameter | Value |
|---|---|
| Total number of signals | 68 |
| Number of electrodes per brain slice | 11 ± 5 |
| Number of ictal events per brain slice | 5 ± 1 |
| Number of interictal events per brain slice | 80 ± 50 |
| Brain region | Parahippocampal cortex |
| Sampling rate | 2 kHz |
| Total recording time (average) | 1800 s |
| Signal-to-noise ratio (SNR) | from 20 ± 10 dB |

**Table 2. Algorithm optimization experiment matrix**. Optimization matrix for the two algorithms. Combinations for each of the parameters of the algorithm and experiments.

| Algorithm | Optimization Parameters |
|---|---|
| ZDensityRODE | Threshold [$\sigma$]: 4, 5, 6 |
| | Delta convolutional filter [seconds]: 3, 4, 5 |
| | Lag [seconds]: 0.125, 0.25, 0.5 |
| AMPD | Threshold [$\sigma$]: 3, 4, 5, 6 |
| | Delta Time Integration [$\sigma$]: 1.5, 2, 2.5 |

## Z-score Density-based Robust Outlier Detection Estimation (ZDensityRODE)

Since the LFP signal in the used dataset primarily consists of baseline activity, then we may presume that any deviation from the signal's trend could indicate an interictal or ictal event. By analyzing the outlier's temporal duration, proximity to other outliers, and magnitude, we could somewhat estimate the type of event it may represent. The ZdensityRODE algorithm utilizes the statistical Z-score test to determine outliers in data trends using the distribution's historic mean and standard deviation. The Z-test is a reliable measure of variability that can be used to identify outliers. By using historical data, the algorithm can adjust to changes in amplitude caused by factors like electrode degradation. Once the outliers are identified through a z-score test, they are stored for later spike density analysis. The density of these events is evaluated using a lookup window that meets specific search criteria and then classified based on their temporal characteristics and density.

This algorithm can be configured using four parameters lag ($\lambda$), threshold ($\tau$), influence ($\iota$), and delta ($\Delta$). The lag parameter determines how many past samples are considered for computing the historic average and standard deviation. A longer lag provides stability against rapid changes but increases memory load, while a shorter lag enables quicker adaptation to new information but increased variability. The threshold parameter sets the z-score level at which a data point is deemed significant, allowing for more precise detections based on signal features like the SNR. The influence parameter (ranging from 0 to 1) controls the weight given to incoming samples when recalculating averages and standard deviations, with 0 disregarding new data and 1 providing high adaptability. Lastly, the delta parameter defines the refractory time between peaks, crucial for pattern classification during the post-processing analysis of peak candidates. This parameter regards the duration of events and their separation in time. The algorithm workflow is depicted in Figure 3. The pseudocode for this algorithm is also provided in Code 1. The algorithm's complexity was assessed, resulting in a $O(n + m)$ complexity, where n is the input data size and m is the number of relevant outliers found in the signal. If the signal contains no outliers or those are very limited, the spike density classifier is much faster as there are fewer events to compare.

The algorithm starts with a training stage, in which during the first $\lambda$ samples, the algorithm does not provide any detection but fills the buffers and calculates the initial trend of the data. Only a notch filter at 50 Hz to remove powerline noise (here, 50 Hz) is used in the signal conditioning stage. During this stage, which lasts as long as the lag time parameter, initial mean ($\mu$), and standard deviation ($\sigma$) buffers are filled (See Fig. 3). Once the algorithm is trained, it starts the operation mode by processing each incoming sample in three steps: (i) a z-score test relative to $\mu$ and $\sigma$, which are estimated values from previous samples in the signal (Eq. 1); (ii) local maxima evaluation by first-derivative test (Eq. 2); (iii) refractory period evaluation.

$$z = \frac{x_n - \mu}{\sigma} \quad (1)$$

$$x_{n-1} < x_n \oplus x_{n+1} < x_n \quad (2)$$

Where $x_n$ is the current sample, $\mu$ is the historic mean and $\sigma$ is the historic standard deviation. Both historic $\mu$ and $\sigma$ are implemented as FIFO, storing the $\lambda$ last samples and recalculating on each sample with the influence parameter depending on the evaluation of the z-score. the $\mu$ and $\sigma$ filters are updated following the expression in Eq. 3. When the z-score exceeds the threshold, it indicates a significant and prominent peak, thus, the historical data is updated based on the influence defined by $\iota$. If the z-score test is negative, implying noise or irrelevant information, the filters $\mu$ and $\sigma$ are updated without considering the influence parameter, that is providing less importance to the current sample. This removes the influence of this value on the adaptation of the threshold.

$$\overline{x_n} \leftarrow \iota \cdot \overline{x_i} - (1 - \iota) \cdot \overline{x_{i-1}} \\ n = 0, 1, \ldots, \lambda \quad (3)$$

When detecting outliers, these are transformed into intervals by observing the density of spikes over a continuous period of time. The number of events is recorded with a timestamp, which will be used to determine the type of event. The time intervals are measured based on the event length, the up-spike and down-spike, and their proximity to the next peak, using a simple incremental counter. If there are one or more peaks that occur within the integration time interval $\Delta$, the detection window is extended until the peaks are separated by a time greater than the refractory period. When this condition is not met, the integration is stopped and the interval is classified. Interval classification is determined by duration, amplitude, and spike density. These characteristics provide information about event frequency and intensity.

## Automatic multiscale-based peak detection (AMPDE)

The scalogram technique has proven to be effective and reliable in identifying peaks, especially in scenarios with noise and various peak shapes [32]. This algorithm uses a multiscale decomposition based on scalograms and various windows, followed by peak identification in each row of the scalogram matrix. As demonstrated by Scholkmann et al. [33], this method shows remarkable performance in detecting relevant events. The algorithm comprises six stages: Signal Conditioning, calculation of Local Maxima Scalogram (LMS), determination of optimal scale, LMS rescaling, peak detection, and estimation of peak density for brain activity pattern classification. A comprehensive block diagram of the algorithm is provided in Figure 4, offering a visual representation of each step within the processing pipeline and input/output (I/O) delineation. The algorithm's pseudocode is also available in Code 2. Its complexity was evaluated, resulting in an $O(n^2)$ time complexity. This is because of the nested loops used in the matrix operations that significantly increase the algorithm's complexity.

Let us now consider a univariate signal, $x(i)$, with N uniformly sampled points, $1 \leq i \leq N$, containing periodic or quasi-periodic peaks. The first step involves preprocessing the input signal to remove the offset and eliminate the power line interference (here, 50 Hz) by applying a notch filter with a quality factor of 90. Then the LMS is computed by analyzing all input values ($x = x_1, x_2, x_3, x_4, ..., x_n$) using a moving window approach. The LMS uses the scalogram for the detection of peaks in the time series. The scalogram is the signal's frequency distribution over time, analogous to the spectrogram. This representation enables the extraction of significant features such as peaks, edges, and patterns. In this work, we are interested in the detection of peaks and this representation provides a trustworthy measure of relevant events and their frequency components. To obtain the LMS matrix, the signal $x$ is first detrended by subtracting the least-squares fit of a straight line from $x$. This operation removes the long-term memory of the signal and avoids adding carried trends into the current portion of the signal where the scalogram has to be computed. Next, the signal is averaged with a moving window average filter with varying lengths. The window length $w_k$ is varied from 2 to L, $w_k = 2k, k = 1, 2, ..., L$, where L is defined as $L = \lceil \frac{N}{2} \rceil - 1$. This is performed for every scale $k$ and for $i = k+2, ..., N-k+1$.

$$m_{k,i} = \begin{cases} 0 & x_{i-1} > x_{i-k-1} \wedge x_{i-1} > x_{i+k-1} \\ r + \alpha & otherwise \end{cases} \quad (4)$$

where r is a uniformly distributed random number in the range [0, 1] and $\alpha$ a constant factor ($\alpha = 1$). Each $m_{k,i}$ element is formed as the value $r + \alpha$ for $i = [1, 2, ..., k+1]$ and for $i = [N-k+2, ..., N]$. The result of performing all those operations for each of the windows is a matrix L x N defined as Eq. 5. Each row in this matrix corresponds to a window length $w_k$. It represents the local maxima scale of the signal for each window. To gather information about the scale-dependent distribution of zeros (and consequently local maxima) in the signal segment, a row-wise summation is performed for each LMS level, as shown in Eq. 6.

$$M = \begin{bmatrix} m_{1,1} & m_{1,2} & \cdots & m_{1,N} \\ m_{2,1} & m_{2,2} & \cdots & m_{2,N} \\ \vdots & \vdots & \ddots & \vdots \\ m_{L,1} & m_{L,2} & \cdots & m_{L,N} \end{bmatrix} \quad (5)$$

$$\gamma_m = \frac{1}{L} \sum_{k=1}^{L} m_{k,m} \quad for\ k \in 1, 2, ... L \quad (4)$$

The vector $[\gamma_1, \gamma_2, \ldots, \gamma_L]$ contains information about the distribution of zeros that is scale dependent, including the local maxima. The next step uses a parameter $\lambda$ to reshape the LMS matrix $M$ by removing all elements $m_{k,i}$ for which $k > \lambda$ holds. It allows one to reshape the LMS matrix leading to the new $\lambda \times N$-matrix ($r_{k,i}$), for $i \in 1, 2, \ldots, N$ and $k \in 1, 2, \ldots, \lambda$. From the matrix, the peak detection is performed by: (i) Calculating the column-wise standard deviation of the matrix according to Eq. 7 and (ii) Finding all indices $i$ that satisfy $\sigma_i = 0$. These values are stored in the vector $\hat{p} = [p_1, p_2, \ldots, p_{\hat{N}}]$ with $\hat{N}$ being the total number of detected peaks of the signal $x$.

$$\sigma_i = \sqrt{\frac{1}{\lambda}\sum_{k=1}^{\lambda}(m_{k,i} - \gamma_k)} \quad for\ i \in 1, 2, \ldots N \tag{7}$$

In the last step, a similar strategy as in the ZdensityRODE is followed to estimate the peak density by considering factors such as amplitude, duration, and proximity to other events. Peaks are searched within a period denoted by $\Delta$, and if one or more peaks are found within that duration, the detection interval is extended until the peaks are separated by a time greater than the refractory time. This process continues as long as the peaks are close and $\Delta$ time is not exceeded..

## Performance metrics

In the context of TSS problems, the evaluation of the performance of an output model relies on its agreement with the reference model. A positive detection corresponds to a sequence of labels that collectively represent a category or type of event in the LFP signal. To compare reference and output annotations, a set-theory-based approach is necessary. To determine a true positive (TP) detection, the intersection between the time intervals associated with the detected event and the reference event has to be found. The degree of overlap between these events settles a robust metric for measuring algorithm performance. The reference and output interval intersection is computed as per Eq. 8. By dividing the coincidence segment by the reference segment, the coincidence ratio is obtained, which quantifies the degree of overlap between the two events (refer to Eq. 9).

$$\text{Coincidence Sequence} = [r_n, r_{n+i}] \cap [y_m, y_{m+j}] \tag{8}$$

$$\text{Coincidence Ratio} = \frac{min(r_{n+i}, y_{m+j}) - max(r_n, y_m)}{r_{n+i} - r_n} \tag{9}$$

Where r and y are reference and output intervals of bounds [n, n+1] and [m, m+j], respectively. Both are defined as vectors since they are a sequence of labels that together constitute events. Here, a TP detection is considered when more than 80% of the output event coincides with the reference events and the length does not exceed 50% of the boundaries of the reference event. Similarly, a False positive (FP) occurs when an event is detected where no reference event is found. False negative falls for a missed detection, either not detected or not sufficient for being classified as TP. In this context, Precision (P +) measures the proportion of true positives over false positives, quantifying how well the output events match the expert's annotations (Eq. 10). Recall (Re) gauges the reliability of the algorithm's actual detections, signifying the fraction of positive patterns correctly classified (Eq. 11). The F-measure, also known as the F1 score, provides a unified score balancing precision and recall. (Eq. 12). We also employ the Jaccard/Tanimoto similarity index (Jaq) to assess the degree of overlap between detected events by comparing reference and output vectors [34], as shown in Eq. 13, where A and B are output and reference intervals from the recorded signal. These metrics provide comprehensive insights into the algorithm's performance.

$$Precision = \frac{TP}{TP+FP} \tag{10}$$

$$Recall = \frac{TP}{TP+FN} \tag{11}$$

$$F\text{-}score = \frac{Precision \times Recall}{Precision + recall} \tag{12}$$

$$Jaccard = \frac{|A \cap B|}{|A \cup B|} \tag{13}$$

A Multi-Criteria Decision Analysis (MCDA) is proposed, to combine precision, recall, Jaccard index, and f-measure to extract a more comprehensive score of the algorithm performance. There is no clear indication as to which metric should be prioritized, nor there is a preferred metric for assessing such kinds of problems. For instance, in [7,18,35], recall is used as a fundamental metric for seizure detection, while in [14, 15, 36–38] precision is used as a key indicator. Sensitivity is also employed by [16, 37–39] or specificity in [38,39]. In this work, we place special emphasis on the accuracy of detecting ictal events. To improve the precision, we incorporate the Jaccard index in addition to interval intersection. Along with precision, recall is also equally important in detecting ictal events and ensuring reliable detection. In the case of interictal events, precision, and recall are given equal importance. The equation presented in Eq. 14 is a scoring function that prioritizes the detection of ictal events over interictal detection. However, the algorithm can be trained using a different metaparameter configuration to match the expected result. This means that the algorithm can be adjusted to detect different types of events' detection depending on the specific needs. During the training stage, the weight assigned to each metric determines how the algorithm is prepared for the online operation mode.

$$\begin{aligned}Score = & \, 0.6 \times Precision_{ictal} + 0.2 \times Jaccard_{ictal} + \\ & 0.1 \times Recall_{ictal} + 0.05 \times Precision_{interictal} + \\ & 0.05 \times Recall_{interictal}\end{aligned} \tag{14}$$

where $Precision_{Ictal}$ is given a weight of 60% and $Jaccard_{interical}$ is given a weight of 20%. The metric $Recall_{Ictal}$ is given 10%, $Precision_{interical}$ is given a 5 % and $Recall_{interical}$ a 5 %.

## Results

LFP signal characteristics vary remarkably among brain regions. This work focuses on LFP recorded from the CTX, which is the primary site of ictal onset in the in vitro model of limbic ictogenesis used in this work [25], consistent with what was reported in several human studies [17,30,40,41]. Fig. 5A illustrates the power spectral density (PSD) plot of epileptiform activity recorded from the entorhinal cortex. The frequency features of these signals were described in previous research, where the meaningful content of the CTX signal spectrum was shown to lie within 0.5 to 30 Hz, with additional relevant components at 350-390 Hz and around 650 Hz [42].

In terms of morphology, ictal events, which are the electrographic correlate of seizures, manifest as sustained high-amplitude, high-frequency discharges. In contrast, interictal epileptiform discharges, occurring between ictal events, are brief electrographic transients of various morphology such as single spikes, spike bursts, or oscillations of low/high frequency [43]. In the in vitro model used for this study, ictal discharges typically last $50 \pm 24$ seconds, whereas interictal events last $1.5 \pm 0.8$ seconds. The baseline section represents the signal's noise floor, denoting the absence of epileptiform events. This state is typified by a low-amplitude, uniformly shaped signal that remains relatively stable over time. In terms of their frequency characteristics, ictal events concentrate their spectral power within the frequency range of 0.5 to 30 Hz. On the other hand, interictal events exhibit diverse natures, leading to variations in their spectral representation across different event types. Notably, High-Frequency Oscillations (HFOs), which are often found within interictal epileptiform discharges [12, 44], span a range of 80 to 500 Hz [44, 45]. These features were used here to classify ictal and interictal events and to distinguish them from the signal baseline. Some indicators for distinguishing those events were (i) morphological features like maximum amplitude, spike slope, absolute mean of windowed events, and successive spike density [7, 14, 46, 47]; (ii) spectral features such as power content across different frequency bands [7, 15, 18, 47–49]; (iii) statistical features like variations in median and variance [7, 14, 22], kurtosis and skewness [46].

In this study, only temporal characteristics were used to differentiate between ictal and interictal events. The primary characteristics utilized were the event amplitude range and duration, as they are simple and easy to calculate. However, other relevant indicators may require additional computation, such as skewness, kurtosis or Shannon entropy. Figure 6 illustrates the significant differences between ictal and interictal events and demonstrates the usefulness of temporal characteristics for classification. This approach was aimed at reducing the algorithms' computational cost. An example of the two algorithms' performance is shown against a sample signal and its reference annotations from the dataset in Fig. 7. Our results demonstrate a high performance of the algorithms in detecting ictal events with high recall and high precision (see Fig 8). To ensure a comprehensive evaluation of the algorithm's performance, a multicriteria decision analysis was conducted considering precision, sensitivity, f1-score, and Jaccard index with different weights for ictal and interictal events depending on which detection is prioritized. Both algorithms consistently achieved high scores, with more than 90% in the global score indicating their robustness and reliability in detecting ictal events, but also interictal events with a precision higher than 50%, but lower sensitivity and Jaccard index. ZdensityRODE achieved a precision of P+ = 93% and recall Re = 93%, with a Jaccard index ($Jaq$) of 86% and F1-score of 91% for ictal event detection. However, for interictal events, it achieved a precision of P+ = 42% and recall of Re = 45%. On the other hand, the AMPDE algorithm achieved a P+ = 96% and Re = 90%, with Jaq = 86% and F1-score of 92% for ictal event detection, and a P+ = 54% and Re = 33% for interictal events. The algorithm was trained to give greater importance to ictal detection, thus reflecting a lower performance for interictal detection. Table 3 provides information about the results for each recording session and for each evaluation metric.

**Table 3. Average evaluation metrics.** This table shows the averaged results for each experimental session in the validation dataset. On the top, are the results for the ZdensityRODE algorithm, and on the bottom, are for the AMPD algorithm.

| brain slice n. | precision Ictal | precision Interictal | recall Ictal | recall Interictal | Jaccard Ictal | Jaccard Interictal | F-score Ictal | F-score Interictal | score |
|---|---|---|---|---|---|---|---|---|---|
| 1 | 0.82 | 0.49 | 1.00 | 0.52 | 0.82 | 0.36 | 0.89 | 0.49 | 0.80 |
| 2 | 1.00 | 0.82 | 0.85 | 0.89 | 0.83 | 0.74 | 0.92 | 0.85 | 0.93 |
| 3 | 1.00 | 0.40 | 1.00 | 0.50 | 1.00 | 0.40 | 1.00 | 0.44 | 0.94 |
| 4 | 1.00 | 0.49 | 0.92 | 0.38 | 0.92 | 0.33 | 0.95 | 0.41 | 0.92 |
| 5 | 1.00 | 0.17 | 0.75 | 0.24 | 0.75 | 0.11 | 0.83 | 0.20 | 0.84 |
| 6 | 0.71 | 0.59 | 1.00 | 0.64 | 0.71 | 0.26 | 0.75 | 0.38 | 0.73 |
| 7 | 1.00 | 0.00 | 1.00 | 0.00 | 1.00 | 0.00 | 1.00 | 0.00 | 0.90 |
| Average | 0.93 | 0.42 | 0.93 | 0.45 | 0.86 | 0.31 | 0.91 | 0.40 | 0.87 |
| brain slice n. | precision Ictal | precision Interictal | recall Ictal | recall Interictal | Jaccard Ictal | Jaccard Interictal | F-score Ictal | F-score Interictal | score |
| 1 | 0.98 | 0.69 | 0.91 | 0.18 | 0.90 | 0.16 | 0.94 | 0.27 | 0.93 |
| 2 | 0.92 | 0.70 | 0.92 | 0.30 | 0.84 | 0.29 | 0.91 | 0.37 | 0.88 |
| 3 | 1.00 | 1.00 | 0.80 | 1.00 | 0.80 | 1.00 | 0.89 | 1.00 | 0.94 |
| 4 | 0.95 | 0.34 | 0.94 | 0.32 | 0.90 | 0.25 | 0.95 | 0.33 | 0.88 |
| 5 | 0.91 | 0.48 | 0.92 | 0.21 | 0.83 | 0.17 | 0.91 | 0.29 | 0.85 |
| 6 | 1.00 | 0.00 | 0.88 | 0.00 | 0.88 | 0.00 | 0.94 | 0.00 | 0.86 |
| 7 | 1.00 | 0.50 | 1.00 | 0.25 | 1.00 | 0.25 | 1.00 | 0.20 | 0.90 |
| Average | 0.96 | 0.54 | 0.90 | 0.33 | 0.86 | 0.31 | 0.92 | 0.38 | 0.89 |

It is relevant to note that the SNR may substantially influence the algorithmic performance. To address this possible issue, we used Pearson correlation analysis to examine the correlation between the algorithm score and the SNR. This analysis resulted in a Pearson's correlation coefficient of r=0.7 within the training dataset and r=0.4 within the validation dataset. This correlation is observable in Fig. 8, where A) and C) represent the corresponding results for each dataset, while B) and D) depicts the SNR ranges for each recording session. The performance of the algorithms indeed correlates with the SNR. This variability is akin to selecting the MEA channels based on their anatomical position and assessing signal quality for SNR threshold compliance. Despite this variability, the algorithm exhibits an outstanding overall performance. Besides, the correct choice of parameters highly determines the algorithm performance as illustrated in Fig. 8 E and F.

## Discussion

We have proposed a new method to detect epileptiform activity by segmenting LFP signals using the ZdensityRODE and AMPDE algorithms. The effectiveness of both algorithms was verified by testing them on epileptiform activity recorded from 4AP-treated rodent brain slices via MEA. Both algorithms demonstrated remarkable performance in detecting ictal activity. It should be noted that these algorithms are capable of detecting interictal events as well, as they perform a time series segmentation task.

To improve epilepsy diagnosis and treatment, several seizure detection algorithms have been proposed. In [18], a magnitude and phase synchronization between electrodes could achieve a 66% true positive rate (TPR). Yoo et al [7], used SVM yielding a TPR of 82.7% for seizure classification. Shoeb et al [50] proposed a machine learning technique through a feature vector for the detection of seizures, achieving a 96% of sensitivity and a 2/24h FPR. Kurtosis, skewness, and coefficient of variation from decimate Discrete Wavelet Decomposition were proposed by [51], achieving a precision of 92.66%. Ronchini et al [38], uses entropy- and-spectrum features for seizure detection algorithm, boasting an accuracy of 97.8%. In [48], seizure detection is done by decimating discrete wavelet decomposition and interquartile range and mean absolute deviation, obtaining a precision of 84.2%, a sensitivity of 98.5% and latency of 1.76 seconds.

Table 4 summarizes the performance and algorithmic approaches for the works mentioned above in a comparative table. Many of those implementations were tailored around the single patient rather than being designed as patient-agnostic approaches. In this regard, there is a great debate on the advantages of one approach over the other [52]. On the one hand, personalized algorithms offer a broad range of adjustable parameters and involve a training process; on the other hand, patient-agnostic approaches employ a fixed-parameter detector without the need for additional training [52, 53]. This decision requires a trade-off between detection accuracy, simplicity, and speed. Following any of those approaches determines the accuracy of the algorithm and its applicability to different patients and scenarios. In this study, we have prioritized the perspective development of a seizure detection method that is not patient-specific but, at the same time, implements minimum levels of customization for improved accuracy.

**Table 4. Comparative analysis of MEA event detection algorithms.**

| Paper | Year | Algorithm approach | Precision [%] | Sensitivity [%] |
|---|---|---|---|---|
| [39] | 2016 | NLSVM | - | 95.7 |
| [18] | 2011 | CORDIC | - | 66 |
| [7] | 2013 | SVM | - | 82.7% |
| [15] | 2018 | Approximate Entropy + FFT | 97.8 | - |
| [14] | 2015 | KNN, SVM, Naive Bayes, logistic regression | 80 | 95.24 |
| [16] | 2020 | SVM | - | 100 |
| [36] | 2018 | CNN | - | 79.2 |
| [38] | 2022 | STDP | 97.8 | 85.4 |
| [50] | 2010 | Feature vector extraction | - | 96 |
| [51] | 2012 | Kurtosis, skewness and coefficient of variation from the decimate DWT | - | 100 |
| [4] | 2013 | Single window count, Multiple window count, spectral entropy | - | 94 |
| [48] | 2014 | Decimate DWT and interquartile range and mean absolute deviation | 84.2 | 98.5 |
| ZdensityRODE | 2023 | Z-score test with memory buffering strategy | (ictal) 93, (interictal) 42 | 93 |
| AMDPEC | 2023 | Multiscale peak detection through DWT | (ictal) 96, (interictal) 54 | 90 |

NLSVM: Non-Linear support vector machines
CORDIC: Coordinate rotation digital computer
FFT: Fast Fourier transform
SVM: Support vector machines
STDP: Spike-Timing Dependent Plasticity
DWT: Discrete wavelet transform
CNN: Convolutional neural networks

The identification of relevant biomarkers in LFP signals poses a significant challenge in the epilepsy field, because of the complex dynamics and different morphologies of ictal and interictal events. Further, the SNR significantly influences the performance of event detection algorithms. In various applications, signal quality assessment techniques improve algorithm performance by preventing unnecessary processing under unfavorable conditions [54, 55]. In this work, a low SNR particularly challenged accurate detection by ZdensityRODE, whereas AMPDE proved robust to additive noise, excelling in analyzing frequency bursts.

## Conclusion

We have introduced a TSS approach as a novel strategy for seizure detection, toward improved seizure control via closed-loop brain stimulation. The detection of pathological biomarkers in LFPs recorded from epileptic patients is fundamental for the timely detection (or better, prediction) of seizures to improve brain stimulation strategies for epilepsy treatment. Furthermore, such electrographic biomarkers can provide supplementary information to advance our understanding of brain states related to epilepsy. ZdensityRODE and AMPDE have proved efficient in biomarkers detection, while using minimal and straightforward computation, as opposed to complex algorithmic approaches, such as neural networks. We believe that not only do these algorithms provide a novel perspective to the field, but they also establish the foundation for a growing subset of algorithms for the classification of epileptiform events and the prediction of seizures.

# Figures

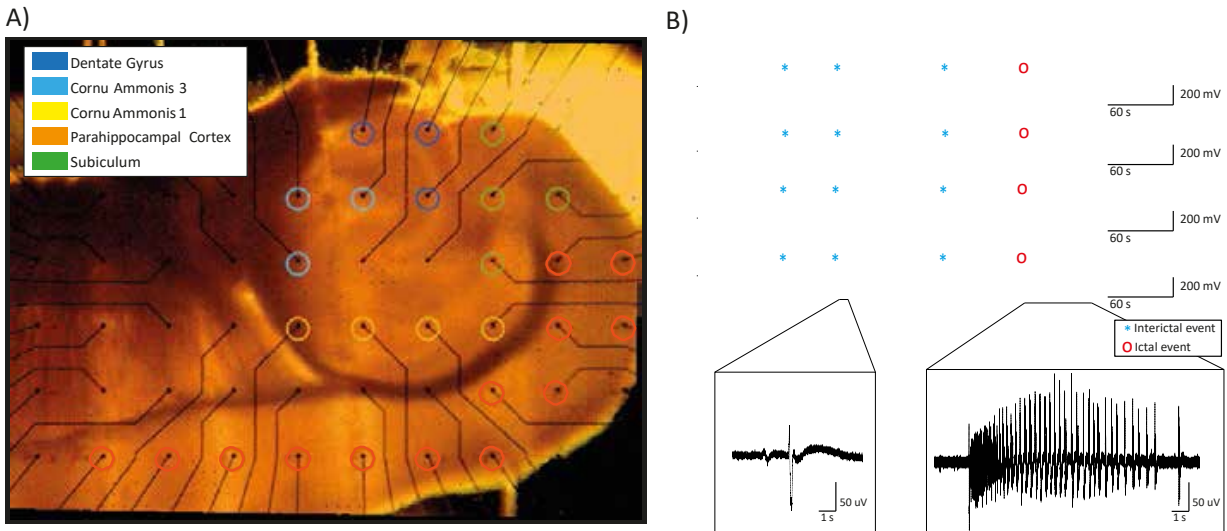

**Fig 1. Epileptiform patterns recorded from 4AP-treated hippocampus-CTX slices via MEA.** (A) Mouse brain slice placed on a planar 6 x 10 MEA grid (grid separation of 500 $\mu m$). Electrodes are grouped by their position in the brain structures comprised within the brain slice preparation. (B) Representative MEA recordings from the electrodes marked as CTX in panel A. The epileptiform pattern comprises both interictal and ictal events. The inserts show representative instances of each event type at an expanded time scale.

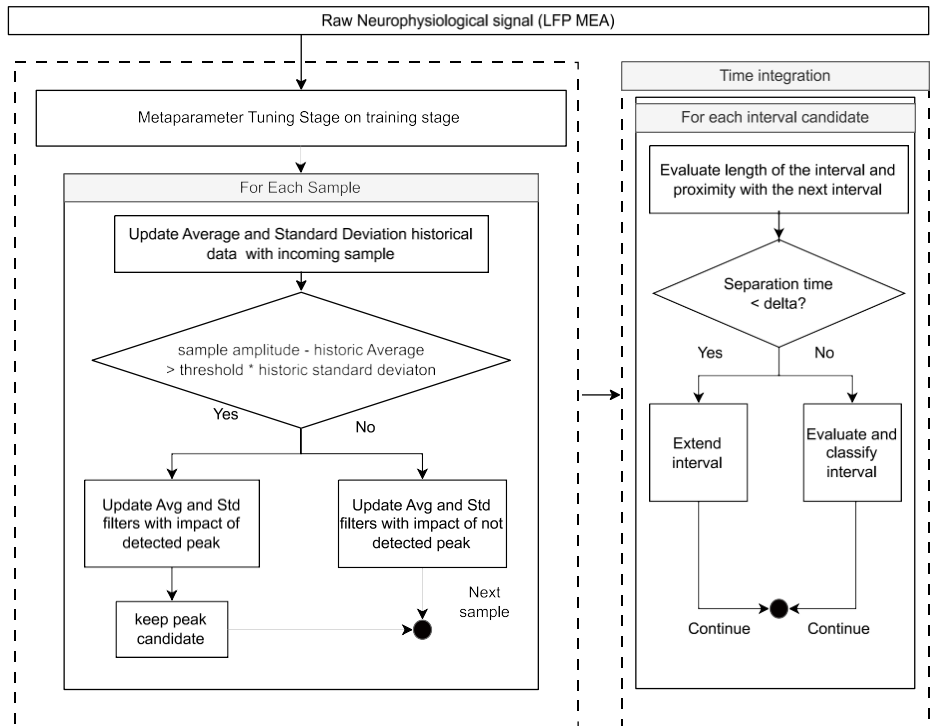

**Fig 3. ZDensityRODE algorithm flowchart.** Block diagram with the workflow of the Z-score Density-based Robust Outlier Detection Estimation algorithm.

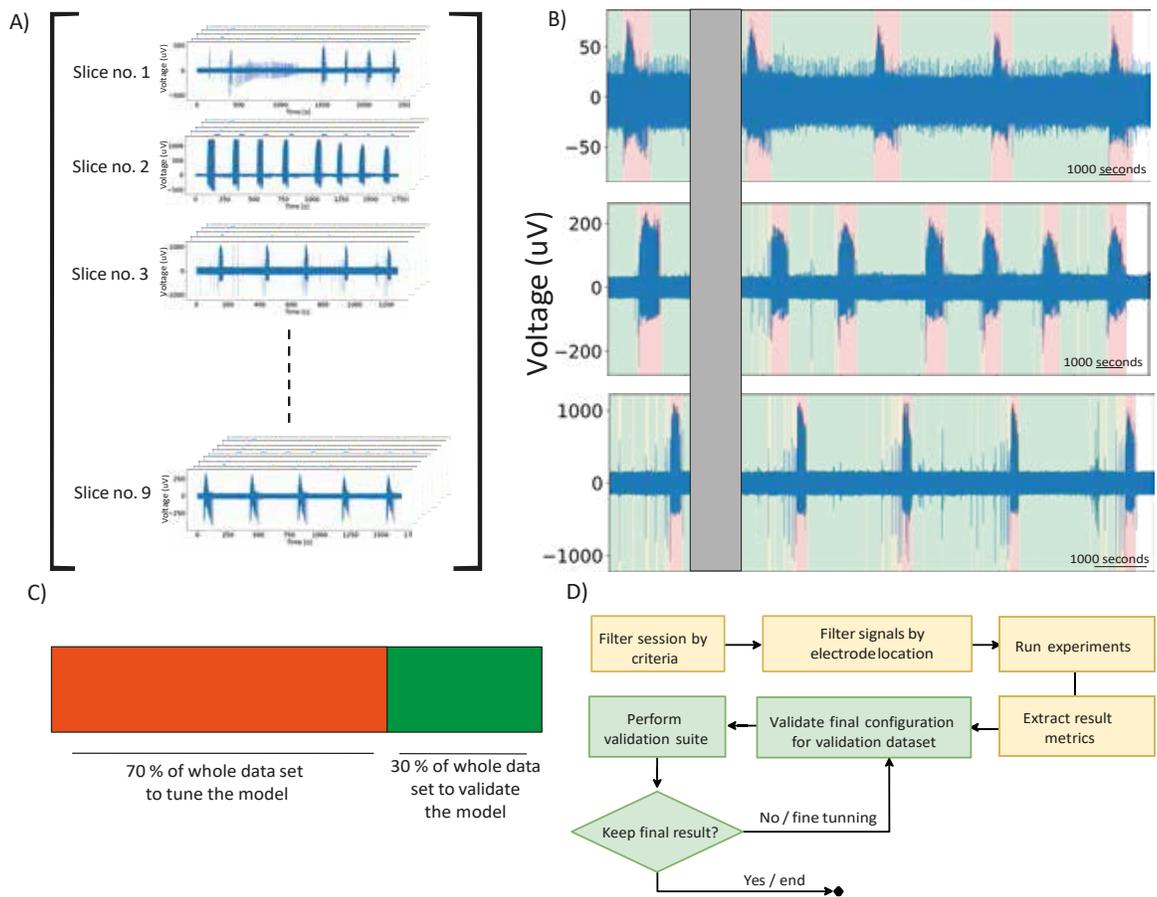

**Fig 2. Experiment design description** A) Compendium of MEA recording datasets. B) Annotated signals showing the three different patterns, highlighted in different colors: baseline (green), interictal (yellow) and ictal (red). C) Test/train split of the dataset. D) Experiment design block diagram with the sequence of stages chosen for the experimental stage.

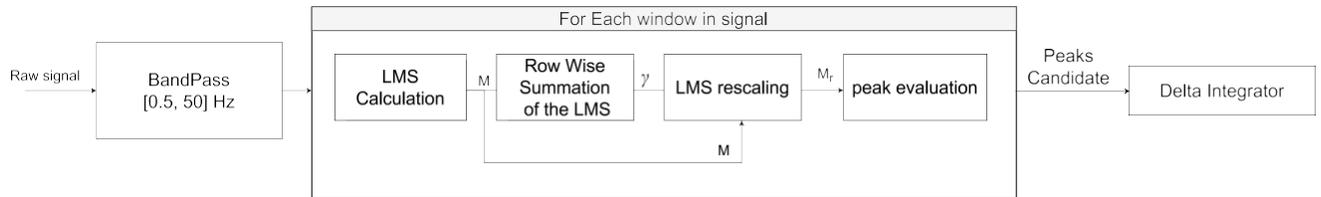

**Fig 4. AMPD enhanced algorithm flowchart.** Block diagram with the workflow of the Automatic multiscale-based peak detection (AMPDE) algorithm. In the first stage, a bandpass filter between 0.5 and 50 Hz is set. Then an LMS calculation is performed for each window in the signal. A summation of each row in the LMS returns the vector γ, which after a rescaling obtains the final vector of peaks candidate, filtered with an evaluation method to identify the peaks on the vector.

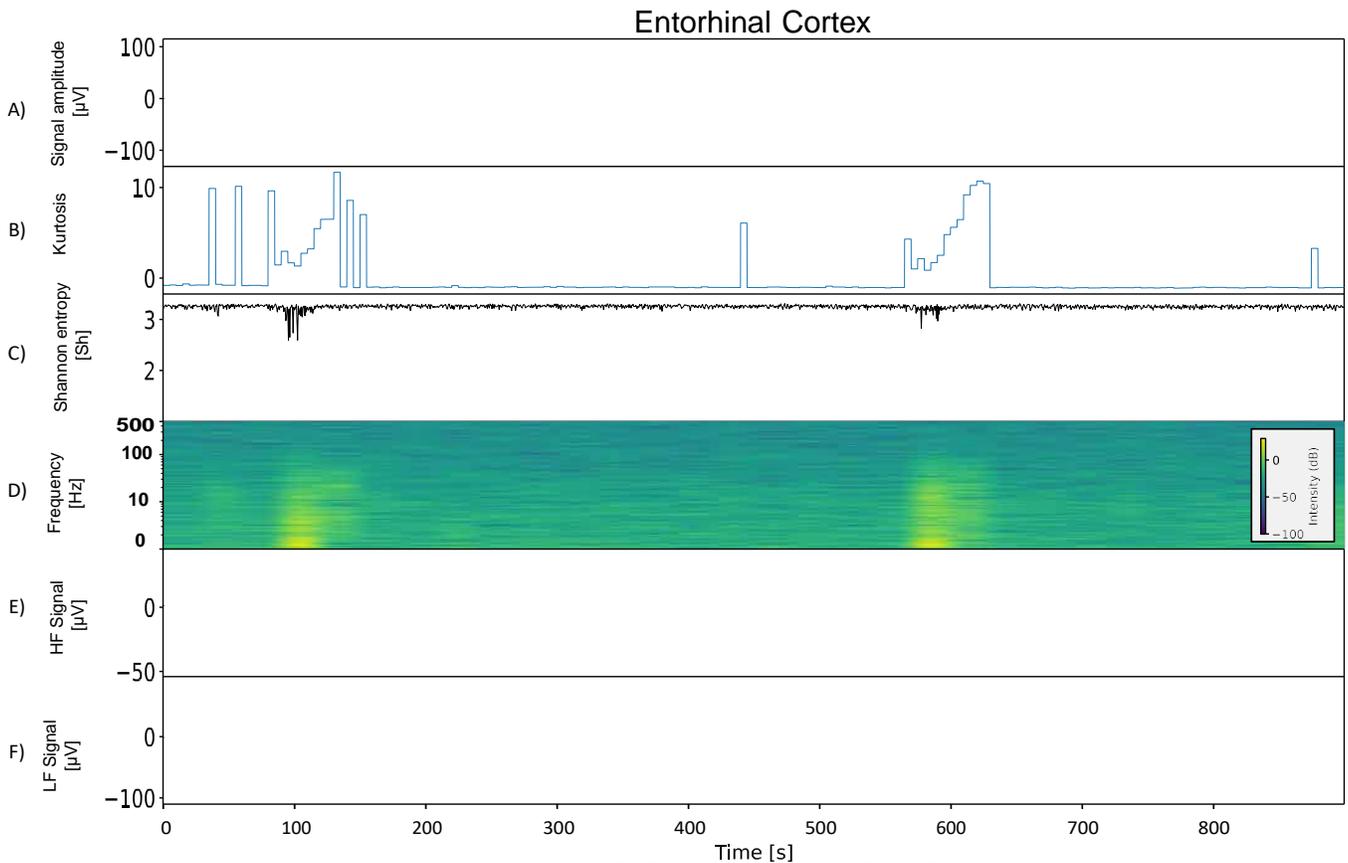

**Fig 5. Features of epileptiform activity recorded from the entorhinal cortex.** A) Raw signal showing recurrent ictal and interictal events. B) Signal Kurtosis computed in 5 s windows. C) Shannon entropy, highlighting the variance and complexity of the signal over time. D) Spectrogram highlighting the wide range of frequency components in epileptiform events. E) Signal after high-pass filtering at $\geq 300 Hz$. F) Signal after low-pass filtering at $\leq 300 Hz$

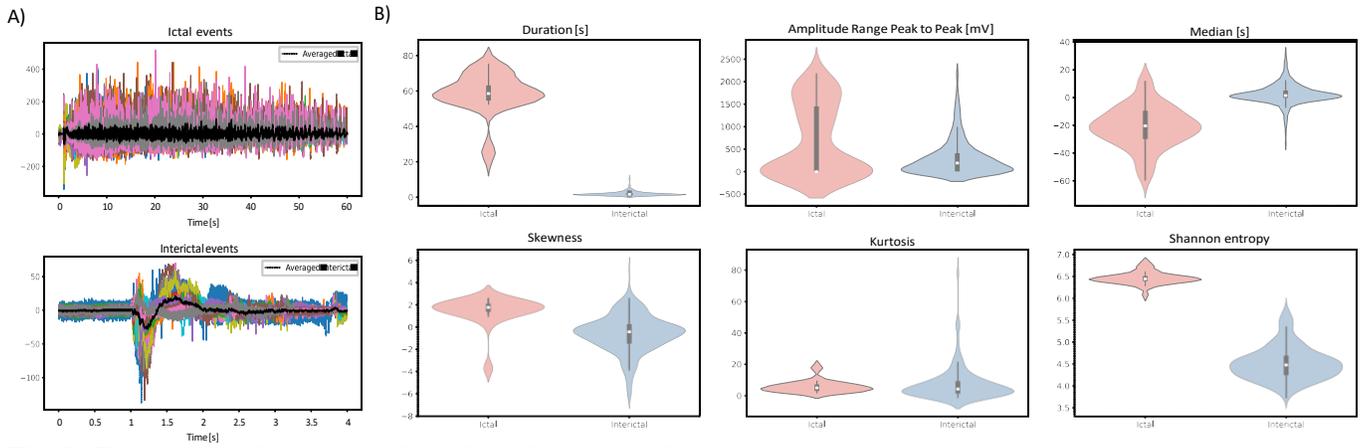

**Fig 6. Distinctive features of ictal and interictal events.** A) Overlayed ictal and ictal discharges and average signal (black). B) Violin plots of a selected number of features, highlighting distinctive characteristics of ictal and interictal events.

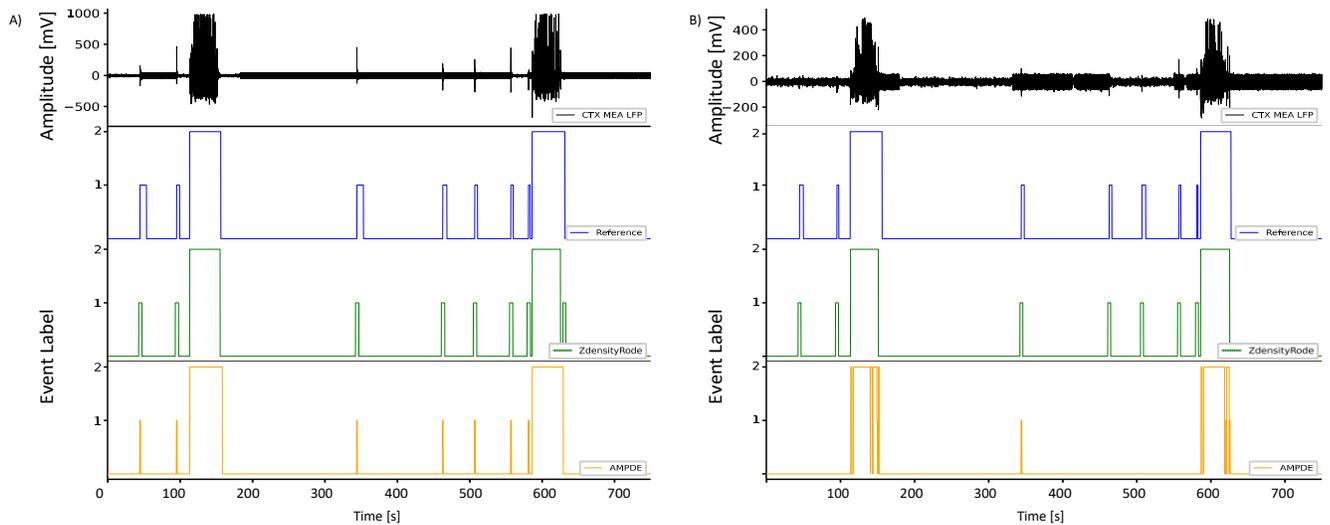

**Fig 7. algorithms Example of segmentation task through ZdensityRODE and AMPDE.** Output comparison for ZdensityRode and AMPDE algorithms against reference annotations for two representative signals, depicting high and low SNR scenarios. A) Representative signal with SNR ≥ 20 dB. B) Representative signal with SNR ≤ 20 dB. (0: baseline, 1: interictal event, 2: ictal event).

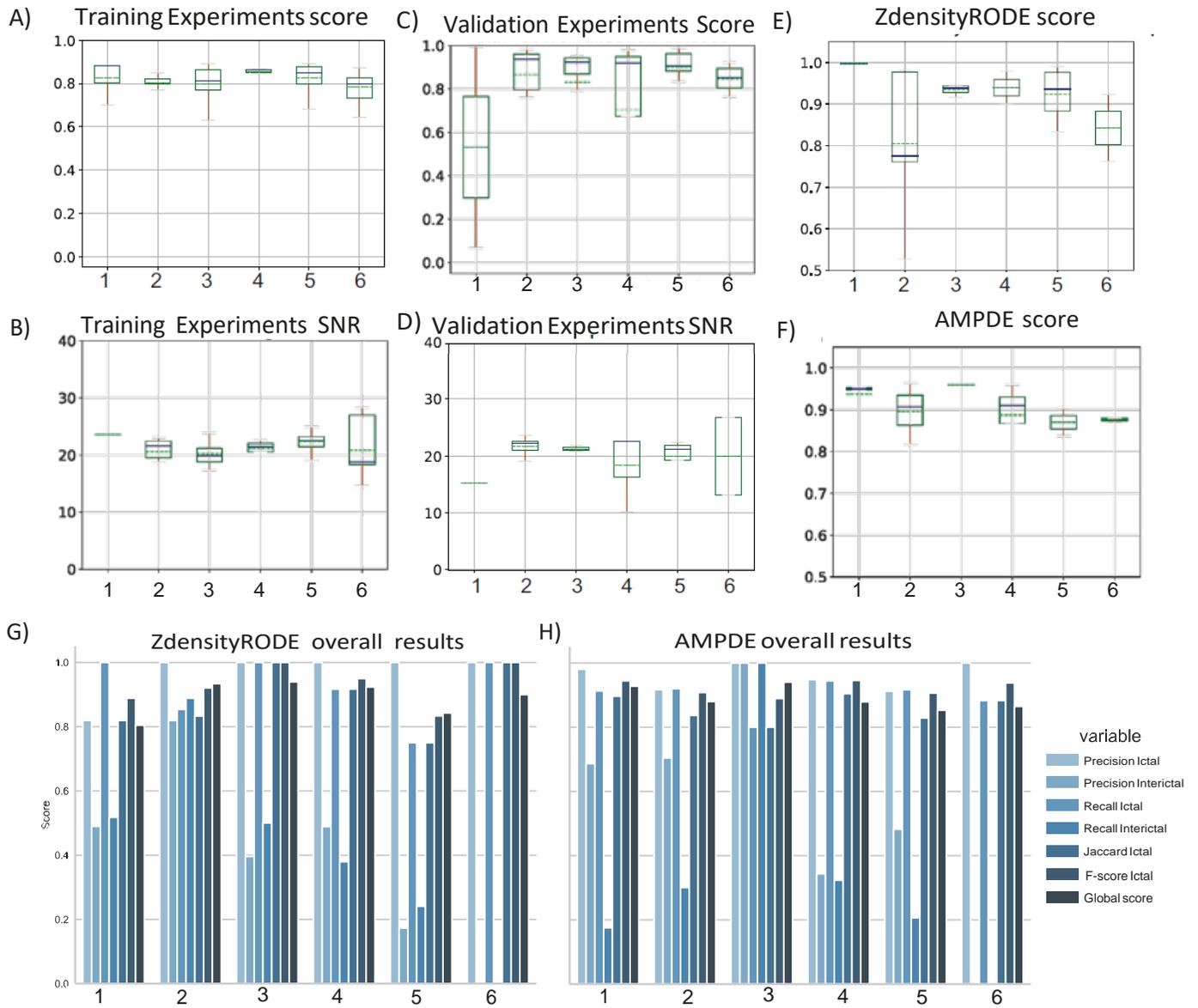

**Fig 8. Experiments results.** A) Distribution of scores for each experimental session in the training stage; B) Distribution of SNR for each experimental session in the training stage; C) Distribution of scores for the validation dataset; D) Distribution of SNR for the validation dataset; E) Distribution of scores for classifier ZdensityRODE; F) Distribution of scores for classifier AMPD; G) Scores for each metric used in the experiment for ZdensityRODE algorithm and H) Scores for each metric used for AMPD algorithm.

**Code 1. Pseudo code for Z-score Density-based Robust Outlier Detection Estimation algorithm.**

---

**Algorithm 1** ZdensityRODE Peak Classification
---
1: **Input:** signal, lag ($\lambda$), influence ($\iota$), threshold ($\tau$), delta ($\Delta$)
2: **Output:** classified regions
3: Define an intermediary signal for filtered input values: Y
4: **for** $x_i$ in signal **do**
5:     **if** Training stage **then** update buffers $\mu$ and $\sigma$ buffers
6:     **else**
7:         **if** $(x_i - \mu_{i-1}) > threshold \cdot \sigma_{i-1}$ **then**
8:             $x_i$ is an outlier
9:             Y := ($\iota \cdot x_i$) + ((1 - $\iota$) * $U_{i-1}$);
10:        **else**U := $x_i$
11:        **endif**
12:     avg := average(Get Chunk of Y for last $\lambda$ samples);
13:     std := Standard deviation(Get Chunk of Y for last $\lambda$ samples);
14:     **for** peak stored in processed window **do**
15:         Account number of peaks in a $\Delta$ window
16:         **if** N peaks in $\Delta$ time-lapse $\geq$ *Threshold A* **then**
17:             Classify interval as A label
18:             **if** *Threshold B* $\leq$ N peaks in $\Delta$ window $\leq$ *Threshold A* **then**
19:                 Classify interval as B label

**Code 2.   Pseudo code for AMPDE algorithm.**

---

**Algorithm 2** Automatic Multiscale-based Peak Detection Enhanced (AMPDE)

---
1: **Input:** Signal ($\mathbf{x}$), Scale range ($\mathbf{k}$), threshold ($\alpha$), delta ($\Delta$)
2: **Output:** Detected peaks $\mathbf{p}$
3: Calculate Linear Detrended Signal $\mathbf{x}'$ by subtracting the least-squares fit of a straight line
4: **for** $k$ in $2^1, 2^2, ..., 2^L$ where $L = \lceil N/2 \rceil - 1$ **do**
5:     **for** $i$ in $k+2, k+3, ..., N-k+1$ **do**
6:         **if** $r + \alpha$ where $r$ is a number $\in [0, 1]$ and $\alpha = 1$ **then**
7:             $m_{k,i} = \begin{cases} 0, & \text{if } x_{i-1} > x_{i-k-1} \land x_{i-1} > x_{i+k-1} \\ r + \alpha, & \text{otherwise} \end{cases}$
8:         **else**
9:             $m_{k,i} = 0$
10:        **end if**
11:     **end for**
12: **end for**
13: Compute Local Maxima Scalogram (LMS) matrix $\mathbf{M}$ using $m_{k,i}$ values
14: Row-wise summation to get $\gamma_k = \sum_{i=1}^{N} m_{k,i}$ for each scale $k$
15: Find global minimum $\lambda = \arg\min_k \gamma_k$
16: Reshape matrix $\mathbf{M_r}$ by removing elements $m_{k,i}$ for $k > \lambda$
17: **for** $i$ in $1, 2, ..., N$ **do**
18:     **for** $j$ in $1, 2, ..., \lambda$ **do**
19:         Calculate $\sigma_i$ using Eq. (7) column-wise standard deviation formula
20:     **end for**
21: **end for**
22: Detect peaks: $p = \{i \mid \sigma_i = 0\}$
23: Peak sorting task
24: **for** each peak, count all consecutive peaks within a delta ($\Delta$) time **do**
25:     **if** next peak exceeds the integration ($\Delta$) window **then**
26:         **if** Number of peaks exceeds classification threshold for ictal **then**
27:             Mark the event as Ictal
28:         **else**
29:             **if** Within interictal kind of events **then**
30:                 Mark the event as Interictal
31:             **end if**
32:         **end if**
33:     **end if**
34: **end for**
        =0

29. Spencer SS, Spencer DD. Entorhinal-hippocampal interactions in medial temporal lobe epilepsy. Epilepsia. 1994;35(4):721–727.

30. Panuccio G, Colombi I, Chiappalone M. Recording and modulation of epileptiform activity in rodent brain slices coupled to microelectrode arrays. JoVE (Journal of Visualized Experiments). 2018;(135):e57548.

31. Caron D, Canal-Alonso A, Panuccio G. Mimicking CA3 Temporal Dynamics Controls Limbic Ictogenesis. Biology. 2022;11(3). doi:10.3390/biology11030371.

32. Bishop SM, Ercole A. Multi-Scale Peak and Trough Detection Optimised for Periodic and Quasi-Periodic Neuroscience Data. Acta Neurochir Suppl. 2018;126:189–195. doi:10.1007/978-3-319-65798-1_39.

33. Scholkmann F, Boss J, Wolf M. An Efficient Algorithm for Automatic Peak Detection in Noisy Periodic and Quasi-Periodic Signals. Algorithms. 2012;5:588–603. doi:10.3390/a5040588.

34. Chung NC, Miasojedow B, Startek M, Gambin A. Jaccard/Tanimoto similarity test and estimation methods for biological presence-absence data. BMC Bioinformatics. 2019;20(Suppl 15):644. doi:10.1186/s12859-019-3118-5.

35. Gregg NM, Marks VS, Sladky V, Lundstrom BN, Klassen B, Messina SA, et al. Anterior nucleus of the thalamus seizure detection in ambulatory humans. Epilepsia. 2021;62(10):e158–e164.

36. Truong ND, Nguyen AD, Kuhlmann L, Bonyadi MR, Yang J, Ippolito S, et al. Convolutional neural networks for seizure prediction using intracranial and scalp electroencephalogram. Neural Networks. 2018;105:104–111. doi:https://doi.org/10.1016/j.neunet.2018.04.018.

37. Ronchini M, Zamani M, Huynh HA, Rezaeiyan Y, Panuccio G, Farkhani H, et al. A CMOS-based neuromorphic device for seizure detection from LFP signals. Journal of Physics D: Applied Physics. 2021;55(1):014001.

38. Ronchini M, Rezaeiyan Y, Zamani M, Panuccio G, Moradi F. NET-TEN: a silicon neuromorphic network for low-latency detection of seizures in local field potentials. Journal of Neural Engineering. 2023;20(3):036002.

39. Bin Altaf MA, Yoo J. A 1.83 $\mu$J/Classification, 8-Channel, Patient-Specific Epileptic Seizure Classification SoC Using a Non-Linear Support Vector Machine. IEEE Transactions on Biomedical Circuits and Systems. 2016;10(1):49–60. doi:10.1109/TBCAS.2014.2386891.

40. Fisher RS, Vickrey BG, Gibson P, Hermann B, Penovich P, Scherer A, et al. The impact of epilepsy from the patient's perspective I. Descriptions and subjective perceptions. Epilepsy Research. 2000;41(1):39–51.

41. Haneef Z, Skrehot HC. Neurostimulation in Generalized Epilepsy: A Systematic Review and Meta-Analysis. Epilepsia. 2023;.

42. Galeote-Checa G, Panuccio G, Linares-Barranco B, Serrano-Gotarredona T. Baseline Features Extraction from Microelectrode Array Recordings in an in vitro model of Acute Seizures using Digital Signal Processing for Electronic Implementation. 2021 IEEE International Conference on Omni-Layer Intelligent Systems, COINS 2021. 2021;doi:10.1109/COINS51742.2021.9524089.
May 1, 2024                                                                                                                                     23/24